\documentclass[english,aps,10pt,english,aps,twocolumn,showpacs,preprintnumbers,amsmath,amssymb,superscriptaddress]{revtex4}
\usepackage[latin9]{inputenc}
\usepackage{graphicx}
\usepackage{float}
\usepackage{amssymb}
\usepackage{esint}
\usepackage{soul}
\usepackage{babel}
\usepackage{hyperref}
\usepackage{placeins}
\usepackage{color}
\usepackage[dvipsnames]{xcolor}

\begin{document}

\title{High frequency torsional motion transduction using optomechanical coupled oscillators}

\author{Hamidreza Kaviani}
\thanks{These authors contributed equally.}
\affiliation{Institute for Quantum Science and Technology, University of Calgary, Calgary, AB, T2N 1N4, Canada}
\affiliation{National Research Council of Canada Nanotechnology Research Centre, Edmonton, AB, T6G 2M9, Canada}

\author{Bishnupada Behera}
\thanks{These authors contributed equally.}
\affiliation{Institute for Quantum Science and Technology, University of Calgary, Calgary, AB, T2N 1N4, Canada}
\affiliation{National Research Council of Canada Nanotechnology Research Centre, Edmonton, AB, T6G 2M9, Canada}

\author{Ghazal Hajisalem}
\affiliation{Institute for Quantum Science and Technology, University of Calgary, Calgary, AB, T2N 1N4, Canada}
\affiliation{National Research Council of Canada Nanotechnology Research Centre, Edmonton, AB, T6G 2M9, Canada}

\author{Gustavo de Oliveira Luiz}
\affiliation{nanoFAB Centre, University of Alberta, Edmonton, AB, T6G 2V4, Canada}

\author{David P. Lake}
\affiliation{Institute for Quantum Science and Technology, University of Calgary, Calgary, AB, T2N 1N4, Canada}
\affiliation{National Research Council of Canada Nanotechnology Research Centre, Edmonton, AB, T6G 2M9, Canada}

\author{Paul E. Barclay}
\email{pbarclay@ucalgary.ca}
\affiliation{Institute for Quantum Science and Technology, University of Calgary, Calgary, AB, T2N 1N4, Canada}
\affiliation{National Research Council of Canada Nanotechnology Research Centre, Edmonton, AB, T6G 2M9, Canada}

\begin{abstract}
Using light to measure an object's motion is central to operating mechanical sensors that probe forces and fields. Cavity optomechanical systems embed mechanical resonators inside optical resonators. This enhances the sensitivity of optomechanical measurements, but only if the mechanical resonator does not spoil the properties of the optical cavity. For example, cavity optomechanical detection of resonators made from optically absorbing materials, or whose geometry does not possess suitable spatial symmetry, is challenging. Here we demonstrate a system that overcomes challenges in measuring high-frequency twisting motion of a nanodisk by converting them to vibrations of a photonic crystal cavity. Optomechanical readout of the cavity then enables measurement of the nanodisk's torsional resonances with sensitivity $5.1\times 10^{-21}-1.2\times 10^{-19}\,\text{Nm}/\sqrt{\text{Hz}}$ for a mechanical frequency range of 5--800 MHz. The nanodisk can be outfitted with magnetic nanostructures or metasurfaces without affecting the optical properties of the cavity, making the system suitable for torque magnetometry and structured light sensing.
\end{abstract}

\maketitle

The enhanced optomechanical interaction  between light and mechanical resonances in nanophotonic cavities allows ultra-precise measurement of force and displacement \cite{ref:aspelmeyer2014co, ref:xia2020opto, ref:sun2012fdc, ref:gavartin2012aho, ref:gomis2014one, ref:ho2015suspended, ref:han2014opto, ref:li2021cavity, ref:liu2021progress}, enabling sensors including accelerometers \cite{ref:krause2012ahm}, ultrasound detectors \cite{ref:basiri2019precision}, mass spectrometers \cite{ref:sansa2020optomechanical, ref:li2012nonlinear}, and atomic force measurement devices \cite{ref:liu2012wcs}. Cavity optomechanical devices can also sensitively detect torque \cite{ref:wu2014ddo, ref:huang2017torsional, ref:he2016optomechanical, ref:kim2013nto, ref:kaviani2020optomechanical}, leading to advances in torque magnetometry \cite{ref:wu2017not, ref:kim2017maf}, spin detection of electrons \cite{ref:kim2016asq}, and measurement of photon spin and orbital angular momentum \cite{ref:he2016optomechanical, ref:kaviani2020optomechanical}. Ideally, mechanical resonators in these systems are integrated directly within the optical cavity to maximize transduction strength: the conversion of mechanical motion to an optical response. However, many of these applications require metallic or magnetic materials that degrade the cavity's properties if they interact with its optical field. For example, torque magnetometers incorporate permalloy nanostructures \cite{ref:wu2017not}, while mechanical transducers of microwave fields incorporate metal  \cite{ref:jiang2020ebp}.  Challenges also arise with resonators whose geometry precludes their integration directly within nanoscale cavities such as photonic crystals and whispering gallery mode resonators whose optical confinement relies upon precisely engineered geometry that can not be perturbed, or whose mechanical resonance spatial symmetry results in vanishing optomechanical coupling to a cavity's optical field. 

Previously, these challenges have been addressed using mechanical resonators whose interaction with the cavity optical field is limited to regions not patterned with lossy material \cite{ref:wu2017not}, or with resonators that only weakly perturb the optical cavity volume \cite{ref:kim2017maf}, enabling sensitive transduction of MHz frequency mechanical resonances. Here we demonstrate a new approach that couples two mechanical resonators--one actuated by an external signal, and one integrated within an optical cavity--allowing individual optimization of the system's actuation and readout properties. We show that through resonant mechanical coupling, a photonic crystal readout cavity can sensitively monitor motion of a high mechanical frequency nanodisk positioned far from the cavity's optical mode. The nanodisk's resonances are suited for actuation by sources of torque commonly encountered in magnetometry \cite{ref:wu2017not} and structured light \cite{ref:kaviani2020optomechanical}, and can be designed to reach frequencies $> 700\,\text{MHz}$, nearly two orders of magnitude larger than previous optomechanical torque sensors \cite{ref:kim2016asq, ref:wu2017not}. The photonic crystal cavity allows sensitive measurement of the nanodisks's mechanical resonances via their resonant mechanical coupling to its mechanical modes. This enables detection of high frequency nanodisk modes, creating a path towards resonant measurement of nanomagnetic dynamics \cite{ref:hajisalem2019tco}, enhancement of the Einstein-de Haas effect \cite{ref:losby2018ram, ref:mori2020einstein}, studies of spin and orbital angular momentum of photons and electrons \cite{ref:kim2016asq, ref:kaviani2020optomechanical, ref:delord2020spin}, and transduction of radiofrequency signals \cite{ref:bagci2014optical, ref:pearson2020radio, ref:hajisalem2019tco}.

\begin{figure}[ht!]
\centering
\includegraphics{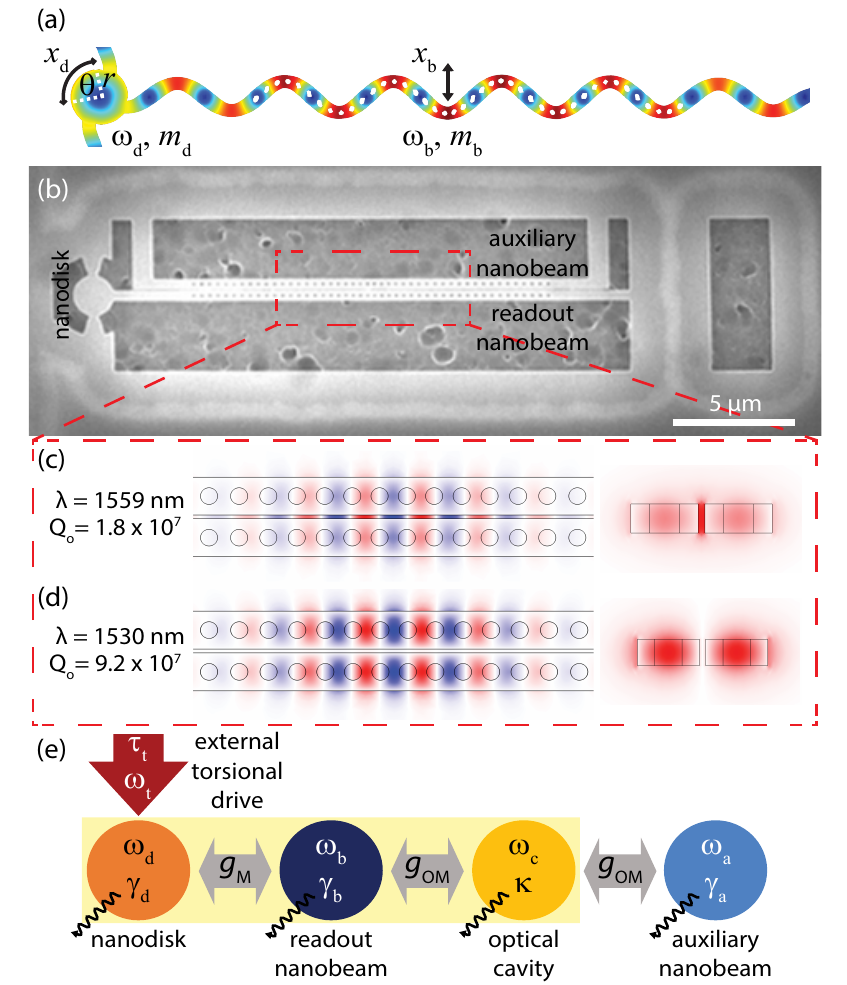}
\caption{(a) Finite element simulation of an in-plane hybridized torsional resonance of the device. Note that the nanodisk's displacement $x_\text{d}$ is related to radius $r$ and angle of rotation $\theta$ of the nanodisk by $x_\text{d} = r\theta$, using the small angle approximation. (b) SEM image of the device. Finite difference time domain simulation of electric field profile of the zipper cavity: (c) fundamental bonded mode with $\lambda = 1559\,\text{nm}$, $Q_\text{o} = 1.8\times10^7$ and (d) fundamental anti-bonded mode with $\lambda = 1530\,\text{nm}$, $Q_\text{o} = 9.2\times10^7$. (e) The torque sensing device (represented inside the yellow box) can be modeled as a coupled oscillator system where the nanodisk and readout nanobeam are mechanically coupled. The readout nanobeam forms one half of a photonic crystal zipper cavity. Motion of both the readout nanobeam and the auxiliary nanobeam of the cavity are optomechanically coupled to the cavity mode. Dynamics of each mode are described by frequency $\omega$ and damping rate $\gamma$. Mechanical coupling between the nanodisk and readout nanobeam resonances is determined by $g_\text{M}$. The cavity optomechanical coupling is described by $g_\text{OM}$.}
\label{fig:Figure 1}
\end{figure}

\begin{figure*}[ht!]
\centering
\includegraphics{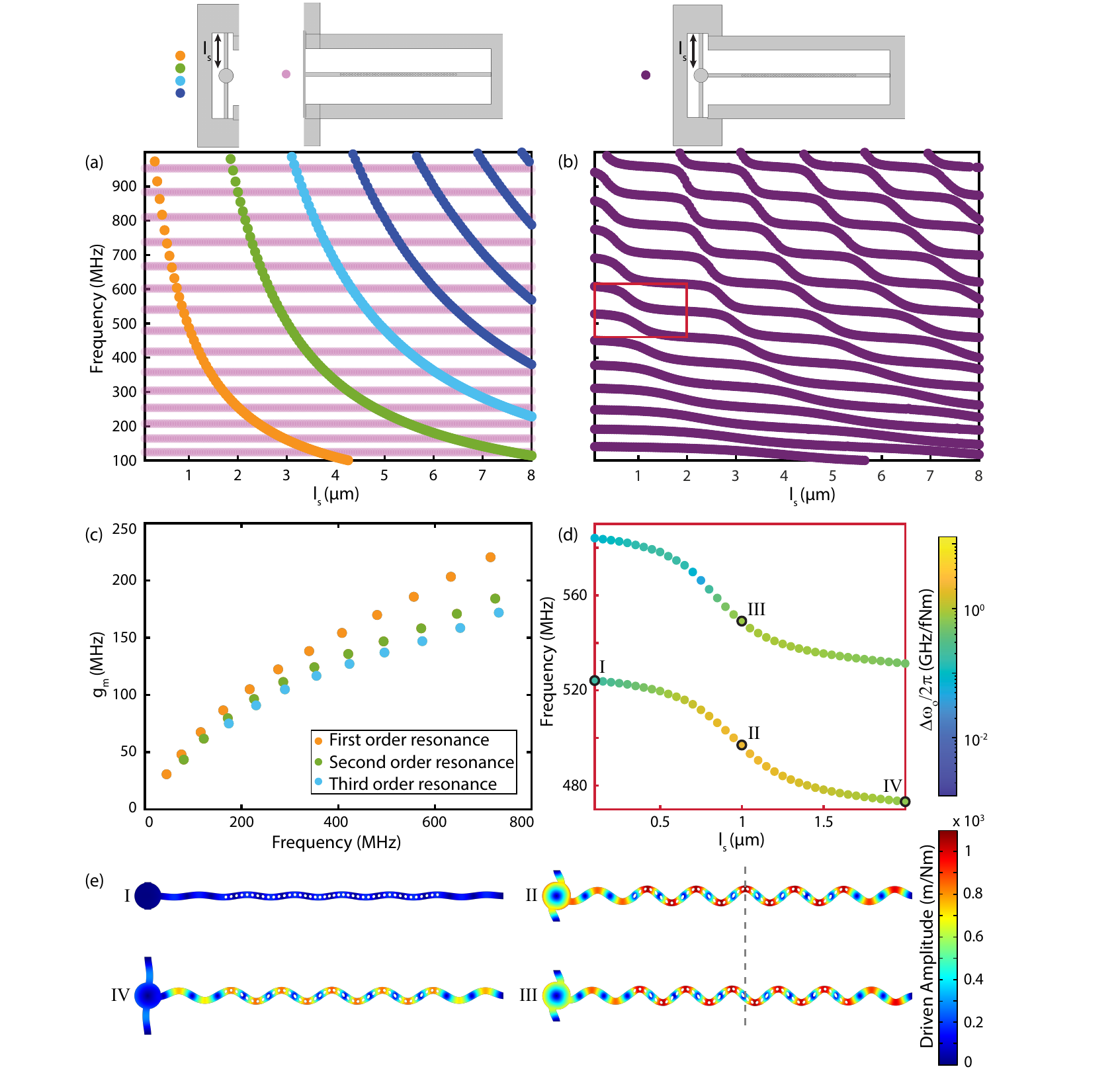}
\caption{(a) Individual simulated mechanical in-plane flexural resonances of uncoupled readout nanobeam (magenta) and torque actuated mechanical resonances of uncoupled nanodisk (orange, green, cyan - for first three resonances respectively, and blue - for rest higher order resonances) as a function of the length of nanodisk's support beams $l_\text{s}$. (b) Simulated eigenfrequency of the in-plane hybridized torsional resonances of the device - coupled nanodisk and readout nanobeam - as a function of $l_\text{s}$. (c) Calculated mechanical coupling strengths $g_\text{M}$ of a typical device for first three harmonics of nanodisk's resonances. (d) A zoom in of the region highlighted by a red box in (b) showing the calculated optomechanical frequency shift per unit applied torque as indicated by the color scale. (e) The displacement profiles of four mechanical resonances with different $l_\text{s}$, labeled I-IV in (d). The dashed grey line suggests the displacement profile at the center of the readout nanobeam, where the optical cavity is localized.}
\label{fig:Figure 2}
\end{figure*}

\begin{figure*}[ht!]
\centering
\includegraphics{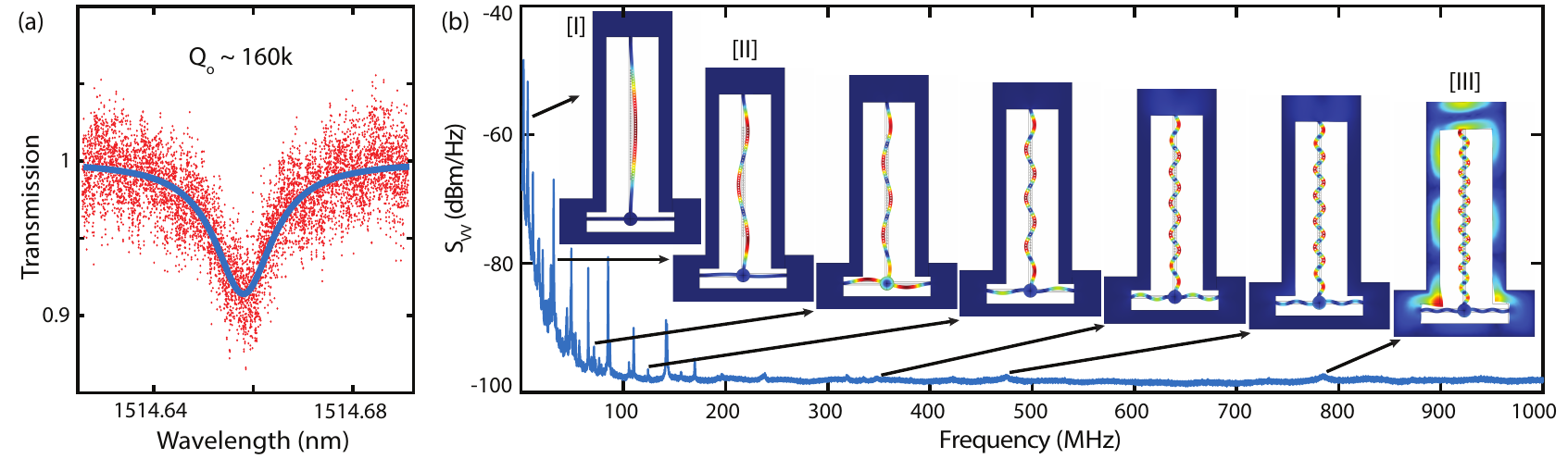}
\caption{(a) Measured optical transmission profile of the torque sensing device. (b) Optomechanical spectrum of the device in ambient conditions. Peaks correspond to thermally driven Brownian motion of the device's mechanical resonances. Peaks associated with torsional modes of the disk are identified. Three specific modes, labeled I-III, are chosen to demonstrate the torque sensitivity of this device, later in Fig.\ \ref{fig:Figure 4}.}
\label{fig:Figure 3}
\end{figure*}

\begin{figure}[h!]
\centering
\includegraphics{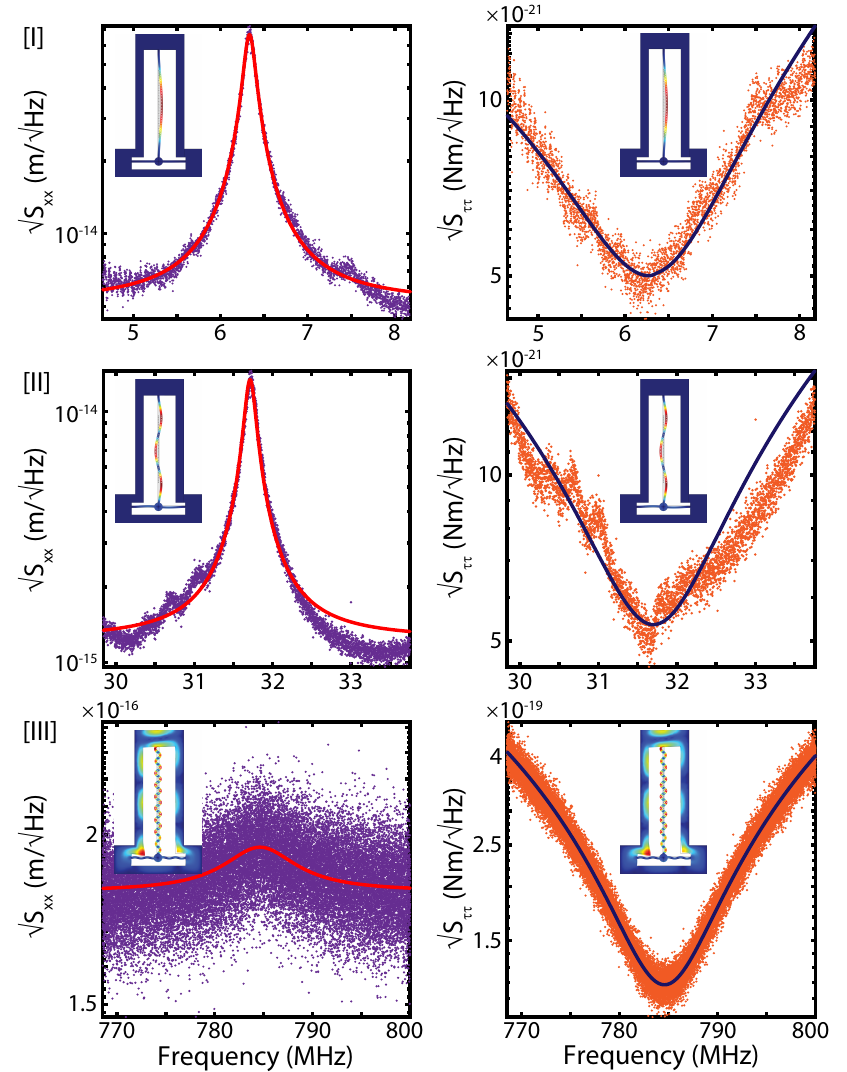}
\caption{Calculated displacement resolution $\sqrt{S_\text{xx}}$ and torque sensitivity $\sqrt{S_{\tau\tau}}$, and their respective fits - red solid line for displacement resolution and dark blue solid line for torque sensitivity, for the three specific modes of interest, labeled [I-III] in Fig.\ \ref{fig:Figure 3}, of the device $l_\text{s}=5.5\,\mu\text{m}$.}
\label{fig:Figure 4}
\end{figure}

The coupled resonator system is illustrated in Fig.\ \ref{fig:Figure 1}(a) and an example of a fabricated device is shown in Fig.\ \ref{fig:Figure 1}(b). A suspended silicon nanodisk anchored to an unpatterned silicon-on-insulator (SOI) chip is connected to a photonic crystal nanobeam that forms one half of an optomechanical zipper cavity \cite{ref:chan2009omd, ref:eichenfield2009apn}. The second half of the zipper cavity is separated by a small gap from the coupled resonators and anchored to the unpatterned region of the chip; it is labeled as the auxiliary nanobeam in Fig.\  \ref{fig:Figure 1}(b). The nanodisk can be mechanically driven by sources of torque ($\tau$), for example from nanomagnetic structures used in torque magnetometry \cite{ref:wu2017not, ref:hajisalem2019tco, ref:losby2018ram}, or by angular momentum transfer from structured optical fields \cite{ref:kaviani2020optomechanical}. Its design addresses limitations of previous photonic crystal torque sensors that detect cantilever-like motion of split-beam cavities \cite{ref:wu2014ddo}, extending both the frequency range and nature of mechanical resonances that can be routinely measured. This is accomplished by resonantly coupling a wide spectrum of high frequency mechanical resonances of the nanodisk to mechanical resonances of the zipper cavity. The zipper cavity combines high optical quality factor $Q_\text{o}$ with large optomechanical coupling. It is formed following the fabrication process in Ref.\ \cite{ref:wu2014ddo}. The suspended photonic crystal nanobeams (width $500\,\text{nm}$, thickness $220\,\text{nm}$) are separated by a gap $70\,\text{nm}$ and patterned with holes of radius 110 nm. The hole spacing in each nanobeam tapers quadratically from $a_\text{o}=400\,\text{nm}$ in the cavity's mirror region to $a_\text{c}=350\,\text{nm}$ at the center of the cavity. The resulting cavity supports fundamental bonded and anti-bonded optical modes shown in Figs.\ \ref{fig:Figure 1}(c) and \ref{fig:Figure 1}(d), respectively, with predicted radiation loss limited $Q_\text{o} > 10^7$ in the 1500 nm wavelength band. Mechanical motion of the nanobeams can change the gap separating them, shifting the cavity mode frequencies. The frequency shift per unit displacement is quantified by optomechanical coupling coefficients $g^{+}_\text{OM}/2\pi \approx 85\,\text{GHz/nm}$ and $g^{-}_\text{OM}/2\pi \approx 2\,\text{GHz/nm}$ between the fundamental in-plane mechanical flexural resonance of the nanobeams and the bonded and anti-bonded optical cavity modes, respectively. The larger optomechanical coupling of the bonded mode arises from its higher field concentration in the gap, making it more susceptible to vibrations of the nanobeams. As a result, we focus on the bonded mode in this study.

In the absence of coupling between the nanodisk and the photonic crystal nanobeam, torque actuated motion of the nanodisk is not transduced by the cavity. However, transduction becomes possible when mechanical coupling is present. Intuitively, this transduction will be enhanced if the nanodisk motion is resonant with a nanobeam mode. This coupled oscillator approach to transduction of the `remote' and nominally dark nanodisk resonances is described by the model illustrated in Fig.\ \ref{fig:Figure 1}(e). Motion of the readout nanobeam connected to the nanodisk is converted to an optical signal through its dispersive cavity optomechanical coupling quantified by $g_\text{OM}$. The dynamics of coupled nanodisk and readout nanobeam are described by equations of motion for the displacement amplitudes $x_\text{d}$ and $x_\text{b}$ of the nanodisk and readout nanobeam resonances, respectively:
\begin{subequations}
\begin{align}
\ddot{x}_\text{d} &= -\omega_\text{d}^{\prime 2} x_\text{d} - \gamma_\text{d} \dot{x}_\text{d}+ \sqrt{\frac{m_\text{b}}{m_\text{d}}}g_\text{M}^2 x_\text{b} + \frac{\tau_\text{t}}{I_\text{d}} f(t)\,,\\
\ddot{x}_\text{b} &= -\omega_\text{b}^{\prime 2} x_\text{b} - \gamma_\text{b} \dot{x}_\text{b}+ \sqrt{\frac{m_\text{d}}{m_\text{b}}}g_\text{M}^2 x_\text{d}\,, 
\end{align}
\label{eq:equations of motion}
\end{subequations}
where $\omega_\text{d}$, $\gamma_\text{d}$, $m_\text{d}$, and $\omega_\text{b}$, $\gamma_\text{b}$, $m_\text{b}$, are the frequency, damping rate, and mass, respectively, of each uncoupled mechanical resonator. Note that  $\omega'_\text{d} = \omega_\text{d} + \sqrt{\frac{m_\text{b}}{m_\text{d}}}g_\text{M}^2$ and $\omega'_\text{b} = \omega_\text{b} + \sqrt{\frac{m_\text{d}}{m_\text{b}}}g_\text{M}^2$ are the modified frequencies, of the nanodisk and the readout nanobeam, respectively, when they are coupled. These equations include an external torsional drive $f(t)$ with an amplitude $\tau_\text{t}$ applied to the nanodisk, whose moment of inertia $I_\text{d}$ is described by an effective mass $m_\text{eff}$ and effective radius $r_\text{eff}$ that is determined from the moment of inertia of the device, which relates the device's angular velocity to its energy, and is defined in \cite{ref:hauer2013general}. Note that for generality an arbitrary drive is used here, where $f(t)$ is a real function. The mechanical susceptibilities $\chi_\text{d,b}(\omega)=(\omega_\text{d,b}^{\prime 2} - \omega^2 - i \gamma_\text{d,b} \omega)^{-1}$ and mechanical coupling $g_\text{M}$ determine how effectively this applied torque is converted into motion of the readout nanobeam. The nanobeam amplitude can be derived by Fourier transforming Eqs. \ref{eq:equations of motion} and solving for $x_\text{b}$ to obtain,
\begin{equation}
x_\text{b}(\omega) = \frac{g_\text{M}^2}{\sqrt{\frac{m_\text{b}}{m_\text{d}}} ((\chi_\text{d}(\omega) \chi_\text{b}(\omega))^{-1} - g_\text{M}^4)} \frac{\tau_\text{t}}{I_\text{d}} f(\omega)\,.
\label{eq:x_rb}
\end{equation}
Equation \ref{eq:x_rb} shows that the resonantly driven transduced amplitude is enhanced, 
\begin{equation}
|x_\text{b}(\omega_\text{m})| = \frac{g_\text{M}^2}{\sqrt{\frac{m_\text{b}}{m_\text{d}}} (\gamma_\text{d}\gamma_\text{b}\omega_\text{m}^2 + g_\text{M}^4)} \frac{\tau_\text{t}}{I_\text{d}}\,,
\end{equation}
when the nanodisk and readout nanobeam are resonant, i.e., ($\omega'_\text{d} = \omega'_\text{b} = \omega_\text{m}$). Here, for simplicity we have chosen the drive amplitude to be unity. The equation also implies that the mechanical mode frequencies become renormalized, as expected for a coupled system, with a frequency difference $\Delta\omega^2 \approx 2 g_\text{M}^2$.

The coupling strength $g_\text{M}$ depends on the device geometry and modes of interest. It can be determined by calculating the mode spectrum of the coupled resonators for varying device parameters such that the nanodisk and nanobeam modes are tuned through resonance. An example of this approach is shown in Fig.\ \ref{fig:Figure 2}. We first show in Fig.\ \ref{fig:Figure 2}(a) the mechanical frequencies of the individual components of the device for varying length $l_\text{s}$ of the support beams anchoring the suspended nanodisk. We see that the nanobeam's in-plane flexural resonance frequencies remain constant, as expected since the nanobeam geometry is not affected by the support length, while the resonance frequencies of the nanodisk's in-plane twisting modes decrease quadratically with increasing $l_\text{s}$. In-plane twisting modes can be efficiently actuated by torque and for simplicity are referred to below as torsional modes.

At certain $l_\text{s}$ values, the mechanical resonances of the uncoupled nanodisk and readout nanobeam are resonant and the resonance frequencies are renormalized by the mode coupling. This is shown in Fig.\ \ref{fig:Figure 2}(b), which plots the full device's simulated mode spectrum for varying $l_\text{s}$. The anti-crossing behaviour at points where the nanodisk and readout nanobeam modes are resonant is a signature of coupling between them. This mechanical coupling is the key feature that makes transduction of torsional actuation of nanodisk to in-plane flexural motion of readout nanobeam possible. Figure \ref{fig:Figure 2}(c) shows $g_\text{M}$ values extracted from the anti-crossing widths as a function of $l_\text{s}$ for the first three harmonics of the nanodisk's twisting resonances. The trend of increasing $g_\text{M}$ with increasing mechanical frequency suggests that higher frequency nanodisk resonances have better overlap with the nanobeam resonances at their clamping point. Similar increases in coupling between modes with increasing frequency are predicted when considering phononic clamping loss in nanomechanical resonators \cite{ref:wilson2008idn}.

Optical detection of nanobeam displacement from torque applied to the nanodisk depends on the optomechanical coupling strength between the hybridized mode and the photonic crystal cavity, the mechanical mode compliance, and the mechanical coupling between the nanodisk and the nanobeam. To predict the efficacy of a given mode to optomechanically detect torque given all of these competing factors, we directly calculate the frequency shift of the cavity mode for a given torque. We achieve this by numerically calculating the mechanical displacement of the full device per unit applied external torque to the nanodisk, from which we then determine the resulting torsional optomechanical frequency shift $\Delta\omega_\text{o} \propto g_\text{M}^2 g_\text{OM}$. This shift, which captures both the device's cavity optomechanical coupling and the mechanical mode coupling, is shown by the color of the points in Fig.\ \ref{fig:Figure 2}(d) for $l_\text{s}$ swept through a typical anti-crossing. The simulated mechanical displacement profiles at several points within the anti-crossing are shown in Fig.\ \ref{fig:Figure 2}(e).
These simulated mechanical modes illustrate the evolution of the mode profiles across a mechanical anti-crossing region demonstrating the strong hybridization of the torsional modes at the center of the anti-crossing. This observation suggests that for torque sensing these strongly hybridized torsional modes are modes of interest. However, to effectively readout these modes, we require a significant non-zero displacement at the center of the nanobeam where the  photonic crystal optical cavity is located. To illustrate that not all hybridized modes satisfy this criteria, we have added a grey dashed line to the center of the mode profiles (II) and (III) in Fig.\ \ref{fig:Figure 2}(e). The lower frequency strongly hybridized mode has an anti-node at its center which results in a stronger interaction with the optical cavity mode, leading to the optomechanical frequency shift $\Delta\omega_\text{o}$ and sensitivity being maximized at the center of the anti-crossing (II). And as we move away from this point towards modes (I) and (IV), we observe a decrease in $\Delta\omega_\text{o}$, as clearly suggested by their smaller amplitudes. On the other hand, the higher frequency strongly hybridized mode has a node at its center, which results in not only zero displacement at the cavity centre, but an odd spatial symmetry of the displacement profile interacting with the cavity mode. This results in a weak interaction with the optical cavity mode, leading to a small $\Delta\omega_\text{o}$ and sensitivity at the center of the anti-crossing (III). More generally, for our cavity design, nanobeam modes with an even number of anti-nodes have small $\Delta\omega_\text{o}$ compared to modes with an odd number of anti-nodes. 

To demonstrate the efficacy of the mode coupling effects discussed above, we experimentally study a range of devices with varying $l_\text{s}$. We optically measure the optomechanically transduced mechanical modes of our fabricated devices using a dimpled optical fiber taper waveguide \cite{ref:michael2007oft} to  evanescently couple light into and out of the cavity. Figure\ \ref{fig:Figure 3}(a) shows the transmission profile of an optical cavity mode from a typical device. The measured mode has $Q_\text{o} \approx 160,000$ and a wavelength in the 1500 nm wavelength range. In a sideband unresolved system where the cavity field instantaneously follows the nanobeam's mechanical motion, the change in the cavity's optical intensity response from an applied torque is $\propto \tau(\omega_\text{t}) \Delta\omega_\text{o} \text{d}H/\text{d}\omega_\text{L}$. Here $\omega_\text{L}$ is the frequency of the probe laser input to the cavity, and $\text{d}H/\text{d}\omega_\text{L}$ is the slope of the cavity mode lineshape, where $H$ is the normalized cavity transmission as a function of laser frequency. Due to the device's high sensitivity arising from its high-$Q_\text{o}$, large $\Delta\omega_\text{o}$, and low mass and associated large mechanical susceptibility (see Eq.\ (2)), thermomechanical driven motion of the mechanical resonances can be observed. Figure \ref{fig:Figure 3}(b) shows a typical mechanical spectrum of a device ($l_\text{s}=5.5\,\mu\text{m}$).

\begin{figure}[ht!]
\centering
\includegraphics{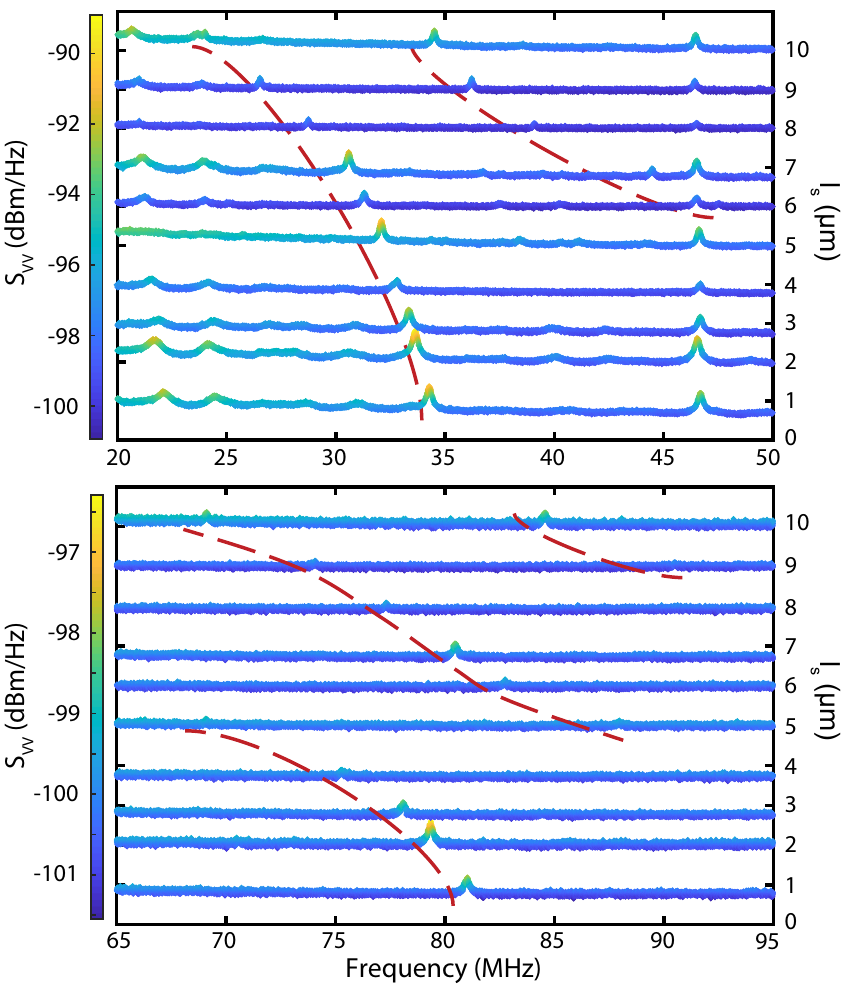}
\caption{Measured mechanical spectrum of the devices for varying support length. The signal strength is color coded. The red dashed line is guide to the eye that follows the moving torsional peaks.}
\label{fig:Figure 5}
\end{figure}

A spectrum analyzer is used to convert the temporal fluctuations in fiber-taper transmission to a power spectral density. Peaks in the spectrum correspond to thermally driven motion of the nanobeam's mechanical resonances. Comparing the frequencies of the peaks in the spectrum with simulated values, we identified the hybridized torsional modes shown in Figure\ \ref{fig:Figure 3}(b) at frequencies that range from $6.3-784.6\,\text{MHz}$. Following the process in \cite{ref:wu2014ddo}, one can extract the displacement resolution $\sqrt{S_\text{xx}}$ and torque sensitivity $\sqrt{S_{\tau\tau}}$ from these measurements as shown in Fig.\ \ref{fig:Figure 4}. As the torque sensitivity $\sqrt{S_{\tau\tau}}$ in our system is limited by thermomechanical noise, it can be extracted from the thermomechanical displacement resolution $\sqrt{S_\text{xx}}$ by, $\sqrt{S_{\tau\tau}} = \frac{m_\text{eff} r_\text{eff}}{|\chi_\text{m}(\omega)|} \sqrt{S_\text{xx}}$. This displacement resolution in our system can be calculated by measuring the signal-to-noise ratio (SNR) of the thermomechanically driven mechanical resonances using the expression $\sqrt{S_\text{xx}} = \sqrt{\frac{4 k_\text{B} T Q_\text{m} / m_\text{eff}\omega_\text{m}^2}{(SNR)}}$, where $k_\text{B}$, $T$, $Q_\text{m}$, are the Boltzmann constant, temperature and mechanical quality factor, respectively. 

To analyze the torque sensitivity, we study three specific modes of interest, labeled [I-III] in the mechanical spectrum of the $l_\text{s}=5.5\,\mu\text{m}$ device shown in Fig.\ \ref{fig:Figure 3}(b), and plot them in Fig.\ \ref{fig:Figure 4}. Based on the fits, we find a torque sensitivity range of $5.1\times 10^{-21}-1.2\times 10^{-19}\,\text{Nm}/\sqrt{\text{Hz}}$ depending on the mechanical mode. In particular, the weakly hybridized torsional mode with frequency $6.3\,\text{MHz}$, which is labeled [I] in Fig.\ \ref{fig:Figure 3}(b) has the highest sensitivity value given above, while the hybridized torsional mode with frequency $784.6\,\text{MHz}$, which is labeled [III] in Fig.\ \ref{fig:Figure 3}(b), is at the lower end of this sensitivity range. The strongly hybridized torsional mode with frequency $31.7\,\text{MHz}$, which is labeled [II] in Fig.\ \ref{fig:Figure 3}(b) also has a high sensitivity value of $5.3\times 10^{-21}\,\text{Nm}/\sqrt{\text{Hz}}$. Though mode [II] is more strongly hybridized than mode [I], because of relatively higher frequency and lower SNR, it has a slightly lesser torque sensitivity than that of mode [I]. Note that these measurements were made at room temperature and in ambient pressure. Operating in vacuum we expect the mechanical quality factors to increase, which will enhance the torque sensitivity \cite{ref:wu2014ddo}. Spectra were typically measured over an 83s acquisition time. Further studies are required to determine how drift or other timescale dependent noise processes will impact the quoted sensitivities.

To further study the hybridization between the nanodisk and nanobeam resonances, the optomechanical spectrum was measured for devices with varying $l_\text{s}$. Figure\ \ref{fig:Figure 5} shows this measurement for $l_\text{s}$ swept from $1\,\mu\text{m}$ to $10\,\mu\text{m}$ in increments of $1\,\mu\text{m}$. As shown by the red trend line, the torsional peaks shift with the expected anti-crossing pattern predicted by the theory discussed in Figs.\ \ref{fig:Figure 2}(a) and \ref{fig:Figure 2}(b). Note that each measurement involves a unique device, and the contribution to the optical transduction related to the optical cavity mode and its coupling to the fiber taper will vary between devices.

\begin{figure}[h]
\centering
\includegraphics{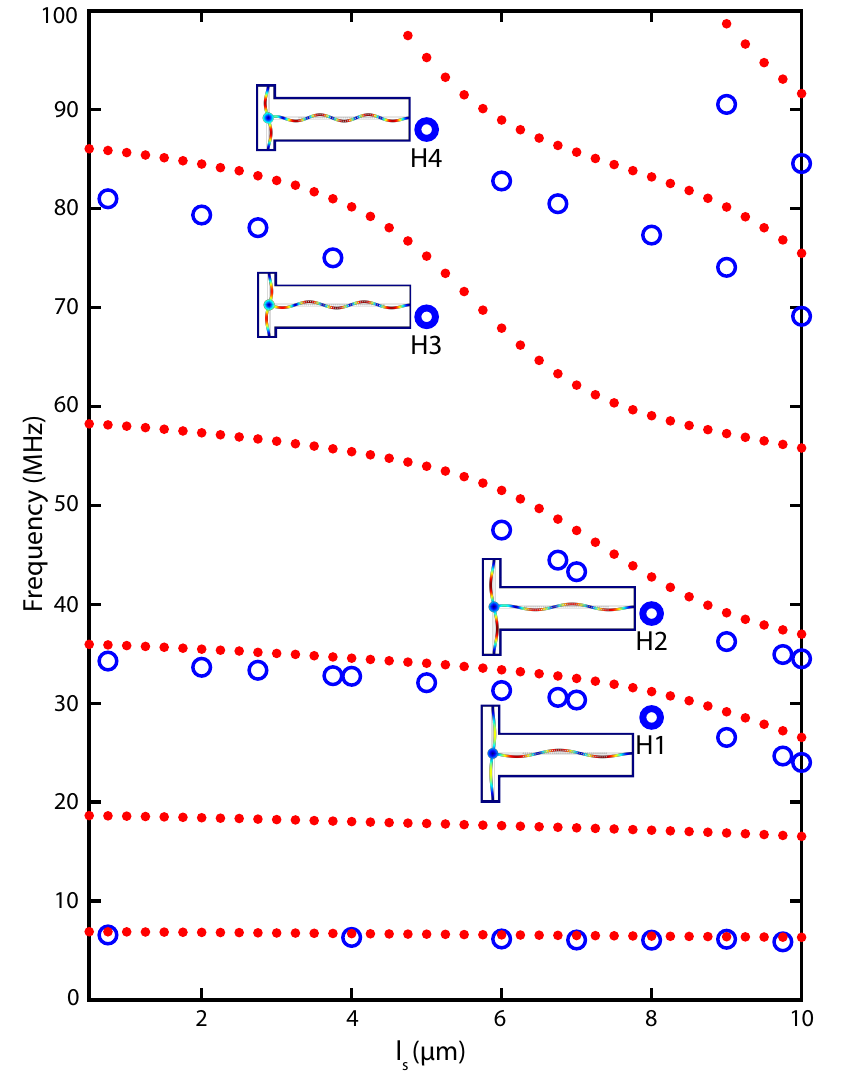}
\caption{Measured mechanical frequencies of the devices with varying support length (blue). The small red dots are simulated eigenfrequencies of the devices.}
\label{fig:Figure 6}
\end{figure}

To compare the measured results with theory more quantitatively, Fig.\ \ref{fig:Figure 6} plots the frequencies of the hybridized torsional resonances measured in Fig.\ \ref{fig:Figure 5} as a function of support length, together with the simulated values. The experimentally observed frequency dependence on support length is in good agreement with the simulated values, with variations most likely attributed to fabrication imperfections. Most crucially, the width of the observed anticrossings closely match the predicted values. It is worth noting that peaks corresponding to nanobeam flexural resonances with an even numbers of anti-nodes do not appear in the spectrum. This can be understood from the fact that optomechanical couplings for these modes is near zero due to their odd symmetry around the center of the optical cavity. 

The device demonstrated here shows how the high sensitivity of nanoscale optomechanical cavities can be harnessed to detect motion of nanomechanical resonators not integrated directly within the optical cavity. By coupling remote and optically dark mechanical resonances to resonators integrated within a photonic crystal nanocavity, mechanical resonances with a wide range of frequencies, mode profiles and symmetries can be sensitively transduced. Since the nanomechanical resonator does not interact directly with the cavity field, it can be decorated with optically lossy metals or magnetic materials. This is of particular interest for advancing the performance and frequency range of optomechanical torque magnetometers used to probe nanomagnetic properties of permalloy \cite{ref:wu2017not}. To this end, the device has demonstrated torque sensitivity of $5.1\times 10^{-21}-1.2\times 10^{-19}\,\text{Nm}/\sqrt{\text{Hz}}$ for a mechanical frequency range of $5-800\,\text{MHz}$. Previous nanoscale cavity optomechanical torque sensing devices \cite{ref:wu2014ddo,ref:kim2013nto,ref:kim2016asq} achieve similar or better sensitivities, but are limited to frequencies below 20 MHz. To the best of our knowledge, the current state-of-the-art in torque sensitivity in an on-chip platform is $\sim 10^{-24}\,\text{Nm}/\sqrt{\text{Hz}}$ at cryogenic temperatures in the $< 20$ MHz mechanical frequency range \cite{ref:kim2016asq}, while the current state-of-the-art in a levitated nanoparticle in vacuum platform is $\sim 10^{-27}\,\text{Nm}/\sqrt{\text{Hz}}$ at room temperature with a rotation frequency $> 5$ GHz \cite{ref:ahn2020ultrasensitive}. The sensitivity of our devices can be further improved by orders of magnitude by performing the experiment in ultra high vacuum and cryogenic temperatures where mechanical quality factors in silicon devices can reach $Q_\text{m} > 10^8$ \cite{ref:maccabe2020nar}. Operation at higher ($> \text{GHz}$) frequencies may be possible using this approach by combining torque actuated mechanical resonators with optomechanical crystal cavities and waveguides \cite{ref:fang2017gnr}.

In the near future, the device's ability to operate at frequencies $>$ 100 MHz will enable high frequency torque magnetometry studies of nanomagnetic systems, opening the door to observation of the Einstein-de Haas effect \cite{ref:losby2018ram, ref:mori2020einstein}, as well as resonant coupling between nanomagnetic spin dynamics and the optomechanical system. Signatures of the Einstein-de Hass effect in driven magnetic spin-mechanic systems become more pronounced at high frequencies \cite{ref:mori2020einstein}. Studies of magnetic resonances typically occur at frequencies above 100 MHz, and one needs to be at a similar or higher mechanical detection bandwidth to study their magnetization dynamics. In cases where the detection bandwidth is much narrower than the rate of angular momentum transfer between the magnetic and mechanical system, the ability to detect a torque will give insight into spin-lattice relaxation processes, which are of fundamental importance to condensed matter physics research. Furthermore, with the integration of the torsional device with magnetic materials it will also become possible to generate effective couplings between light and magnetic spins, with applications in development of hybrid quantum systems \cite{ref:wallquist2009hqd}.

\begin{acknowledgments}
The authors acknowledge helpful discussion with M.\ Mitchell, P.\ Parsa, D.\ Sukachev, M.\ Freeman  and J.\ Losby. This work was supported by the Natural Sciences and Engineering Research Council Discovery Grant and Strategic Partnership Grant programs, the Alberta Innovates Strategic Research Project program, the Canadian Foundation for Innovation, and the National Research Council Nanotechnology Initiative. 
\end{acknowledgments}

\bibliography{nano_bib}

\end{document}